\documentclass{article}
\usepackage{spconf,amsmath,graphicx}
\usepackage{bm,amssymb,algorithm,algorithmic,epstopdf,epsfig,url}
\usepackage{array}
\usepackage[utf8x]{inputenc} 

\def\b#1{{\mathbf #1}}
\def\mrm#1{{\mathrm{{#1}}}}

\def\argmin#1{\underset{{#1}}{\operatorname{argmin}}}

\title{Block randomized optimization for adaptive hypergraph learning}

\name{George Karantaidis \qquad Ioannis Sarridis  \qquad Constantine Kotropoulos \thanks{Acknowledgement: This research has been co-financed by the European Union and Greek national funds through the Operational Program Competitiveness, Entrepreneurship and Innovation, under the call RESEARCH - CREATE - INNOVATE (project code: T1EDK-02474). }}

\address{Department of Informatics, Aristotle University of Thessaloniki\\
Thessaloniki, 54124, Greece\\
Email: gkarantai@csd.auth.gr \quad isarridis@csd.auth.gr \quad  costas@aiia.csd.auth.gr}

\begin{document}
\ninept
\maketitle
\begin{abstract}
The high-order relations between the content in social media sharing platforms are frequently modeled by a hypergraph. Either hypergraph Laplacian matrix or the adjacency matrix is a big matrix. Randomized algorithms are used for low-rank factorizations in order to approximately decompose and eventually invert such big matrices fast. Here, block randomized Singular Value Decomposition (SVD) via subspace iteration is integrated within adaptive hypergraph weight estimation for image tagging, as a first approach. Specifically, creating low-rank submatrices along the main diagonal by tessellation permits fast matrix inversions via randomized SVD. Moreover, a second approach is proposed for solving the linear system in the optimization problem of hypergraph learning by employing the conjugate gradient method. Both proposed approaches achieve high accuracy in image tagging measured by $F_{1}$ score and succeed to reduce the computational requirements of adaptive hypergraph weight estimation.
\end{abstract}
\begin{keywords}
Adaptive hypergraph learning, Randomized algorithms, Block randomized singular value decomposition, Conjugate gradient method.
\end{keywords}
\section{Introduction}
\label{sec:intro}

Social media platforms store huge amount of multimedia content daily and  encourage users to provide descriptions and tags about it. This perpetual, dynamic procedure peaks with users sharing the content among the community. As a consequence, ever increasing data are stored every second in companies' servers. Handling these data becomes a very important task. The era of big data has motivated research towards integrating and developing new methods to cope with large volumes. Large-scale matrix analysis requires huge amounts of resources (i.e., time, memory). 

Randomized algorithms for such large-scale matrices are widely used in order to derive approximate low-rank matrix factorizations. They exploit the structure of matrices to provide partial decompositions. For example, in hypergraph learning, one has to cope with a large hypergraph Laplacian or adjacency matrix, posing difficulties in direct matrix inversions or Singular Value Decomposition (SVD) computation. In addition, large-scale problems may involve inaccurate or missing values, which cannot be effectively dealt by classical methods \cite{Kotropoulos:15}. The main, underlying idea of randomized techniques for low-rank matrix approximations is to seek for a subspace that captures most of the action of the raw matrix \cite{halko2011finding}. Afterwards, common deterministic approaches can be applied. It is shown through extensive experiments and detailed error analysis that randomized algorithms for low-rank approximations are often more robust, accurate, and faster than the classical methods, such as the direct SVD \cite{halko2011finding}.

Recently, a low-rank approximation of large-scale matrices was proposed using a sparse orthogonal transformation matrix for reducing data dimension \cite{Hatamirad2018}. In \cite{martinsson2011randomized}, an efficient randomized scheme for approximate matrix decomposition with SVD was presented. An efficient probabilistic scheme with finite probability of failure was introduced \cite{liberty2007randomized}. A fast deterministic method to solve the high dimensional low-rank approximation was proposed \cite{teng2018fast}. There, a randomized algorithm was adopted and more information was exploited thanks to a sparse subspace embedding. Recently, a low-rank approximation of sparse matrices based on Lower-Upper (LU) factorization was presented \cite{grigori2018low}. Column and row permutations were adopted, searching for an optimal trade-off between speed and accuracy. In \cite{oh2018fast}, a fast and accurate method of a closed-form proximal operator for solving nuclear norm minimization and its weighted alternative was proposed. This method reduced significantly the computational cost due to avoiding a direct computation of SVD. An approximate basis, capturing the range of the input matrix, was formed by its compressed edition.

Hypergraphs are widely used in order to model high-order relations between heterogeneous vertices in an abundance of scientific fields. Solid theoretical basis is offered by hypergraphs in a variety of applications in multimedia search, data mining, mathematics etc. \cite{Kotropoulos:15}. The mainspring in social media applications is to grasp efficiently users' interaction. Image tagging is a typical example of users' interaction. The context information (e.g., user friendships and geo-tags) included in image tagging can be expressed in terms of high-order relations. Therefore, hypergraphs are suitable for modeling this kind of relations, where heterogeneous vertices (i.e., users, user social groups, tags, geo-tags, and images) are linked with hyperedges \cite{Yu:12a,Xu:12}. The influence of each hyperedge is of crucial importance and can be assessed by properly estimating its weight
\cite{Gao:13,Pliakos:15}. The adaptive gradient descent hypergraph learning scheme  presented in \cite{7471862}, is extended by implementing block randomized SVD in optimization step to reduce time requirements. In this first appoach, even though randomized methods are suitable for low-rank matrices, matrix tesselation enables the application of randomized SVD to full-rank matrices met in the optimization problems addressed. Creating rank deficient blocks in the main diagonal, allows low-rank submatrices to be inverted, reducing execution time through block randomized SVD \cite{hackbusch1999sparse}. Moreover, a second application for image tagging is proposed that benefits from the conjugate gradient method employed in the solution of the associated optimization problem. Using these two approaches, we can exploit the benefits of the adaptive hypergraph learning scheme even when a large number of images are used as input. These two approaches are the novel contributions of this work, enabling the application of adaptive hypergraph learning to even larger hypergraphs than that considered here. It is demonstrated that both approaches achieve accurate image tagging measured by the $F_{1}$ measure as in \cite{Pliakos:14tag} and succeed to reduce drastically the computational time requirements.

The outline of the paper is as follows. In Section \ref{sec:hypmodel}, the general hypergraph model is introduced and the adaptive hyperedge weight updating model is presented. In Section \ref{sec:RA}, the proposed approaches for adaptive hypergraph learning optimization are detailed. The dataset is briefly described in Section \ref{sec:experim} and the experiments conducted are demonstrated. Conclusions are drawn in Section \ref{sec:concl} .

\section{Adaptive hyperedge weight updating model}
\label{sec:hypmodel}

Let $|\cdot|$ denote set cardinality, $ || \cdot ||$ be the $l_{2}$-norm of a vector, and $\mathbf{I}$ denote the identity matrix of compatible dimensions. A hypergraph $G(V,E,w)$ captures high-order relationships in social media, where $V$ is the set of vertices, $E$ is the set of hyperedges and $w()$ is real-valued function assigning weights in hyperedges to indicate the relative importance of each high-order relationship captured by the hyperedge. The vertex set $V$ is made by concatenating sets of objects of different type (users, social groups, geo-tags, tags, images). Let $m = |V|$ and $n=|E|$. The incidence matrix $\mathbf{H} \in \mathbb{R}^{m \times n}$ has elements $H(v,e) = 1 $ if $v\in e$ and $0$ otherwise. The following degrees are defined: $\delta(v) = \sum_{e \in E} w(e) H(v,e) $ and  $\delta(e)=\sum_{v \in V}H(v,e)$, which appear in the main diagonal of the vertex degree diagonal matrix $\mathbf{D}_{v} \in \mathbb{R}^{m \times m}$ and hyperedge degree diagonal matrix  $\mathbf{D}_{v} \in \mathbb{R}^{n \times n}$, respectively. Let $\mathbf{W} \in \mathbb{R}^{n \times n}$ be the diagonal matrix having as elements the hyperedge weights $\mathbf{w}(e)$, $e \in E$ in its main diagonal. To measure the degree of similarity of any pair of vertices, one has to compute matrix $\mathbf{A} \in \mathbb{R}^{m \times m}$ \cite{7471862}: 

\begin{equation} \label{eq:A}
\mathbf{A} =\mathbf{D}_{v}^{-1/2}\mathbf{H}\mathbf{W}\mathbf{D}_e^{-1}\mathbf{H}^{\top}\mathbf{D}_{v}^{-1/2}.
\end{equation}
Let $\mathbf{L}=\mathbf{I}-\mathbf{A} \in \mathbb{R}^{m \times m}$ be Zhou's normalized Laplacian of the hypergraph \cite{Zhou:06}. To cluster the vertices of the hypergraph, one has to minimize

\begin{equation} \label{eq:Tau}
T(\mathbf{f})=\mathbf{f}^{\top} \mathbf{L} \mathbf{f}
 \end{equation}
with respect to (w.r.t.) $\mathbf{f} \in \mathbb{R}^{m}$. Highly connected vertices are meant to have equal values in the optimal ranking vector $\mathbf{f}^{*}$  \cite{Agarwal:06}. So, there is high possibility two images to be very similar, if they share a number of common tags above a specified threshold. Both $\mathbf{A}$ and $\mathbf{L}$ are frequently sparse matrices. The clustering optimization problem (\ref{eq:Tau}) can be treated as a ranking problem by importing the $l_{2}$ regularization norm in order to force the ranking vector $\mathbf{f}$ to be as much as possible equal to a query vector specified by a user  \cite{Zhou:03,Bu:10}. Let $\mathbf{X} = \mathbf{I} - \dfrac{1}{1+\theta}\mathbf{A} \in \mathbb{R}^{m \times m}$, where $\theta$ is regularization parameter. Given $\mathbf{W}$, the best ranking vector minimizing (\ref{eq:Tau}), is given by \cite{Zhou:03,Bu:10}:

\begin{equation} \label{eq:Busolution}
\mathbf{f}^{*} = \frac{\theta}{1+\theta} \mathbf{X}^{-1} \mathbf{y}.
\end{equation}
The next step is to update $\mathbf{W} \in \mathbb{R}^{n \times n}$, using the steepest descent method, as proposed in \cite{7471862}. Let $\mathbf{w} = \left(w_1,\,w_2,\,\dots,\,w_n\right)^{\top}$ be the vector formed by the diagonal elements of $\mathbf{W}$. Meaningful constraints on $\mathbf{w}$ are $\b{1}_{n}^{{\top}} \b{w} =1$ and $\b{w} \geq \b{0}$. Let $P(\b{w})=\b{f}^{{\top}} \b{L} \b{f} + \kappa ||\b{w}||^{2}$, where $\kappa$ is a positive regularization parameter. When $\mathbf{f}$ is fixed, the optimization problem w.r.t. $\mathbf{w}$ is defined as

\begin{equation}\label{eq:optprob}
\argmin{\b{w}} \; P(\b{w})
\;\;\;\;\;\;\;\;\;\; \mbox{s.t. $\b{1}_{n}^{\top}\b{w}=1$ and $\b{w} \geq \b{0}$}.
\end{equation}
where the abbreviation s.t. stands for subject to. The Lagrangian of (\ref{eq:optprob}) is given by $S = P + \sum_{j=1}^{\wp} c_{j} \, G_{j},$ where  $c_{j}$,  $j=1, 2, \ldots, \wp$ are the Lagrange multipliers and  active constraints $G_{j}$ are given by:

\begin{equation}
\label{eq:lagrangemultis}
  G_{j}:  \left\{ \begin{array}{ll}
     \b{1}_{n}^{\top} \: \b{w}-1=0  & \mbox{for $j=1$}  \\
     w_{\nu_j-1} = 0           & \mbox{for $j>1$.} \\
                          \end{array}  \right.
\end{equation}
The equality constraint $G_1$ is always active. The weights of the remaining active constraints stay $0$, where 
$\nu_{j}-1 \in [1,n]$ is an index of a hyperedge weight. The steepest descent rule for updating the weights, $\mathbf{w}^{\mrm{new}} =\mathbf{w}^{\mrm{old}} - \mu \nabla S$, was studied in detail in \cite{7471862}. Here, we are interested in the derivation of fast algorithms for solving (\ref{eq:Busolution}) by exploiting either block randomized SVD of $\mathbf{X}$ or the conjugate gradient method to solve

\begin{equation}
\label{eq:conjLinearSys}
  \mathbf{X}\, \mathbf{f} = \frac{\theta}{1+\theta}\, \mathbf{y}.
\end{equation}

\section{Block Randomized SVD and Conjugate Gradient method for optimization}
\label{sec:RA}

Randomized matrix approximations are performed in two stages. The first stage comprises random sampling in order to find a lower-dimensional subspace which captures the most of the action of $\mathbf{X}\in \mathbb{R}^{m \times n}$. To do so a column orthonormal matrix $\mathbf{S} \in \mathbb{R}^{m \times l}$ should be computed such that

\begin{equation}
 {|| \mathbf{X}-\mathbf{S} \mathbf{S}^{\top} \mathbf{X}||}_{F} \leqslant \epsilon
\end{equation}
where ${||\cdot||}_{F}$ denotes the Frobenius norm of a matrix. If $\mathrm{rank(\mathbf{X})} = k \leqslant m$, an oversampling parameter $\pi$ is chosen, such that $l=k+\pi$. Frequently, $l$ is set equal to $2k$. An $m\times l$ matrix $\mathbf{\Omega}$ whose elements stem from a Gaussian distribution with zero mean and unit variance is essential for the randomized subspace iteration. In this case, a small oversampling parameter $\pi$ equal to $5$ or $10$ yields accurate results  \cite{halko2011finding}.

 The second stage of the randomized SVD algorithm consists of the approximate SVD factorization of matrix $\mathbf{B}$
\begin{equation}
\mathbf{B} = \mathbf{S}^{\top} \mathbf{X}
\end{equation}
i.e., $\mathbf{B} = \mathbf{\tilde{U}} \mathbf{\Sigma} \mathbf{V}^{\top}$. The final solution for $\mathbf{U}$ is given by multiplying the approximate $\mathbf{\tilde{U}}$ with the basis $\mathbf{S}$. The detailed randomized SVD via subspace iteration algorithm is summarized in Algorithm \ref{alg:ra}.

The results of the approximate SVD solution should satisfy
\begin{equation}
 || \b{X}-\b{U} \b{\Sigma} \b{V}^{\top}|| \leqslant \epsilon.
\end{equation}
Steps $2$-$3$ of Algorithm \ref{alg:ra} can be replaced by forming:
\begin{equation}\label{classicRA}
\textbf{Y} = (\mathbf{X}\mathbf{X}^{\top})^{\pi} \mathbf{X}\,\mathbf{\Omega}.
\end{equation}
 To construct matrix $\mathbf{S}$, whose columns form an orthonormal basis for the range of $\mathbf{Y}$, one may employ QR factorization. (\ref{classicRA}) is more sensitive to round-off errors and thus, when high accuracy is required, Algorithm \ref{alg:ra} is preferable. Its superiority is grounded on the orthonormalization step included.

To apply randomized SVD to $\mathbf{X}$, $\mathbf{X}$ has to be low-rank. In our case, $\mathbf{X}$ defined as $\mathbf{X} = \mathbf{I} - \dfrac{1}{1+\theta}\mathbf{A}$ is full-rank. To exploit the benefits of randomized SVD, we partition $\mathbf{X}$, as suggested in \cite{hackbusch1999sparse}, as follows


\begin{algorithm}
\caption{Randomized Singular Value Decomposition for low-rank matrix approximation via subspace iteration}
\textbf{Inputs:} An $m \times m$ matrix $\mathbf{X}$, an integer $\pi$, and an integer number $l$ \newline
\textbf{Output:} Approximate factorization of $\mathbf{X}$, where $\mathbf{U}$ and $\mathbf{V}$ are orthonormal  and $\mathbf{\Sigma}$ is non-negative and diagonal, containing the eigenvalues of $ \mathbf{X}$. 
\begin{itemize}
\item[1] Create a $m\times l$ matrix  $\mathbf{\Omega}$  whose elements are independent identically
distributed Gaussian random variables with zero mean and unit
variance.
\item[2] Form $\mathbf{Y}_{0} = \mathbf{X} \mathbf{\Omega}$ and compute its QR factorization $\mathbf{Y}_{0} = \mathbf{S}_{0} \mathbf{R}_{0}$
\begin{algorithmic}
\FOR{$i=1,2,\,\dots,\,\pi$}\STATE Form $ \tilde{\mathbf{Y}}_{i} = \mathbf{X}^{\top} \mathbf{S}_{i-1}$ and compute its QR factorization $ \tilde{\mathbf{Y}}_{i} = \tilde{\mathbf{S}}_{i} \tilde{\mathbf{R}}_{i}$.\\
Form $\mathbf{Y}_{i} = \mathbf{X} \tilde{\mathbf{S}}_{i}$ and compute its QR factorization $\mathbf{Y}_{i} = \mathbf{S}_{i} \mathbf{R}_{i}$
\ENDFOR
\end{algorithmic}
\item[3] $\mathbf{S} = \mathbf{S}_{\pi}$
\item[4] $\mathbf{B} = \mathbf{S}^{\top} \mathbf{X}$
\item[5] Compute SVD of the matrix $\mathbf{B} = \mathbf{\tilde{U}} \mathbf{\Sigma} \mathbf{V}^{\top}$
\item[6] Set $\mathbf{U} = \mathbf{S} \mathbf{\tilde{U}}$
\end{itemize}
\label{alg:ra}
\end{algorithm}

\begin{equation}
\mathbf{X} = \begin{bmatrix} 
\mathbf{X}_{11} & \mathbf{X}_{12} \\
\mathbf{X}_{21} & \mathbf{X} _{22}
\end{bmatrix}
\end{equation}
where $\mathbf{X}_{11}$ and $\mathbf{X}_{22}$ are low-rank submatrices.
The inverse of matrix $\mathbf{X}$ is given by \cite{hackbusch1999sparse}:

\begin{equation}
\bigg[\begin{array}{cc}
\mathbf{X}_{11} & \mathbf{X}_{12}\\
\mathbf{X}_{21} & \mathbf{X}_{22}\\
\end{array}
\bigg]^{-1} = \bigg[\begin{matrix} \mathbf{Z}_{1}^{-1} & -\mathbf{X}_{11}^{-1}\mathbf{X}_{12}\mathbf{Z}_{2}^{-1} \\
-\mathbf{Z}_{2}^{-1}\mathbf{X}_{21}\mathbf{X}_{11}^{-1} & \mathbf{Z}_{2}^{-1} \end{matrix}\bigg]
\end{equation}
where
\begin{eqnarray}
\mathbf{Z}_{1}&=& \mathbf{X}_{11}-\mathbf{X}_{12}\mathbf{X}_{22}^{-1}\mathbf{X}_{21}\\
\mathbf{Z}_{2}&=&\mathbf{X}_{22}- \mathbf{X}_{21} \mathbf{X}_{11}^{-1}\mathbf{X}_{12}
\end{eqnarray}
To compute $\mathbf{Z}_{1}^{-1}$ and $\mathbf{Z}_{2}^{-1}$ as well as $\mathbf{X}_{11}^{-1}$ and $\mathbf{X}_{22}^{-1}$ randomized SVD is applied.

For further reducing the time requirements of inversions, nested tessellations of submatrices created along the main diagonal are employed until the minimum rank of the created submatrices reaches a minimum value of 50. Regarding the second update of $\mathbf{X}$ after having completed the alternating update of $\mathbf{W}$, the minimum rank of the diagonal submatrices of $\mathbf{X}$ is increased to $500$. In this iteration, matrix $\mathbf{X}$ gets even sparser due to the fact that many of its elements are set to zero after steepest descent. Thus, block randomized SVD enables further computational time reduction in solving (\ref{eq:Busolution}).

Having derived the approximate SVD of $ \mathbf{X} $, i.e., $ \mathbf{X} = \mathbf{U} \mathbf{\Sigma} \mathbf{V}^{\top}$

\begin{equation}
\mathbf{X}^{-1} = \mathbf{V} \mathbf{\Sigma}^{-1} \mathbf{U}^{\top}
\end{equation}
with $\mathbf{\Sigma}^{-1} = \mathrm{diag(\dfrac{1}{\sigma_{1}},\dots,  \dfrac{1}{\sigma_{m}})}$ is non-negative, diagonal and $\mathbf{U}$, $\mathbf{V}$ are orthonormal.

In the second approach, to solve (\ref{eq:conjLinearSys}) the conjugate gradient method is employed, which has found to yield further time reduction. Iteratively, the ranking vector $\mathbf{f}_{i}$, the residual $\mathbf{r}_{i}$ and the search direction $\mathbf{p}_{i}$ are updated starting from an initial $\mathbf{f}_{0}$. Then, the initial residual vector is given by $\mathbf{r}_{0} = \frac{\theta}{1+\theta}\, \mathbf{y} - \mathbf{X}\,\mathbf{f}_{0}$ and the search direction is initiated as $\mathbf{p}_{0}=\mathbf{r}_{0}$. For $i\geq 1$, the updated vectors are given by \cite{barrett1994templates}:

\begin{eqnarray}
\mathbf{f}_i&=& \mathbf{f}_{i-1}+\alpha_i \,\mathbf{p}_{i-1}\\
\mathbf{r}_i&=&\mathbf{r}_{i-1}-\alpha_{i} \,\mathbf{X} \,\mathbf{p}_{i-1}\\
\mathbf{p}_i&=&\b{r}_{i}+\beta_{i}\,\mathbf{p}_{i-1}
\end{eqnarray}
where 
\begin{equation} \alpha_i=\frac{\mathbf{r}_{i-1}^{\top}\,\mathbf{r}_{i-1}}{\mathbf{p}_{i-1}^{\top}\, \mathbf{X} \,\mathbf{p}_{i-1}}
\end{equation}
and
\begin{equation}
\beta_i=\frac{\mathbf{r}_{i}^{\top}\, \mathbf{r}_{i}}{\mathbf{r}_{i-1}^{\top} \,\mathbf{r}_{i-1}}.
\end{equation}

\section{Dataset description and Experiments}
\label{sec:experim}

The same dataset used in \cite{Pliakos:15,7471862} is employed here, retaining the same experimental setup. It contains a large amount of Greek places of interest along with valuable information related to them. In particular, geotagged photos, both indoor and outdoor, are accompanied with auxiliary information, such as id, title, owner, latitude, longitude, tags, and views. Only images having many views captured by users, who frequently upload many images, are retained. Images with many views are assumed to depict worth seeing landmarks which have attracted the interest of active users, demonstrating dense social relations (e.g., possessing many friends, participating in many social groups). The aforementioned information was crawled from social media sharing platforms, e.g. \textit{Flickr}. Our interest is limited to groups that have at least $5$ members (i.e., image owners). The specific cardinalities are summarized in Table~\ref{tab:objects}. A $64$-bit operating system with an Intel(R) Core(TM) $i7-4771$ CPU at $3.5$ GHz and $16$ GB RAM was used in the experiments conducted.

For each picture a set of tags was created and a vocabulary was generated, including the times each tag appeared. Hierarchical clustering was applied afterwards. Detailed information about the dataset, the pre-processing procedure, and hypergraph construction can be found in \cite{Pliakos:15, 7471862}.

\begin{table}[tb]
\caption{Dataset objects, notations, and counts.}
\label{tab:objects}
\begin{center}
    \begin{tabular}{ | l | l | r |}
    \hline
    Object & Notation & Count\\ \hline
    Images & $Im$ & $1292$\\
    Users & $U$ & $440$\\
    User Groups & $Gr$ & $1644$\\
    Geo-tags & $Geo$ & $125$\\
    Tags & $Ta$ & $2366$\\
    \hline
    \end{tabular}
\end{center}
\end{table}

The query vector $\b{y}$ is initialized by setting the entry
corresponding to the test image $Im$ and its owner $U$ to $1$. The tags $Ta$ connected to this image are set equal to $A(Im,Ta)$. The objects corresponding to $Gr$ and $Geo$ associated to the image owner $U$ are set equal to $A(U,Gr)$ and $A(U,Geo)$, respectively. The query vector $\b{y}$ has a length of $5867$ elements. During testing, the tags contained in the test set were not included in the training procedure. To allow comparisons with the previous works \cite{Pliakos:14tag, 7471862}, the same metric has been adopted, i.e., the $F_1$ measure. The $F_1$ measure is calculated at various ranking positions and four of them are included in the detailed comparison (i.e., $F_{1}$@1, $F_{1}$@2, $F_{1}$@5, $F_{1}$@10). Here, our main goal is to demonstrate computational time reduction while maintaining the same performance with the methods in \cite{Pliakos:14tag, 7471862}.

ITH-HWEG stands for the adaptive weight estimation method with steepest descent using a fixed adaptation step \cite{7471862}. Let BR-ITH-HWEG refer to the adaptive weight estimation, using steepest descent and exploiting the block randomized SVD via subspace iteration in (\ref{eq:Busolution}). Similarly, CG-ITH-HWEG refers to solving (\ref{eq:conjLinearSys}) by employing the conjugate gradient method. Let ITH denote the method proposed in \cite{Pliakos:14tag}. We have employed in ITH either block randomized SVD, yielding BR-ITH or the conjugate gradient method, obtaining CG-ITH. The main difference between ITH-WHEG and ITH lies in the fact that in ITH the weights are fixed and (\ref{eq:A}) is calculated only once.

\begin{figure}
  \includegraphics[width=\linewidth]{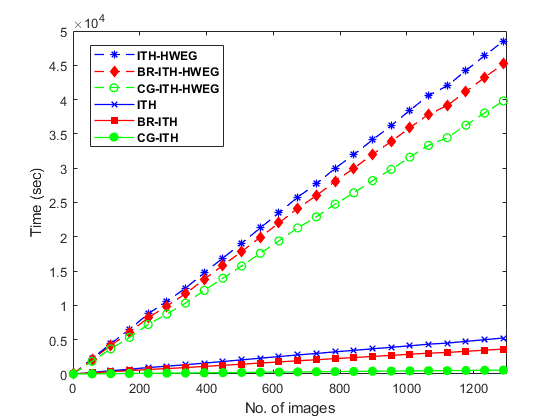}
  \caption{Total time requirements for each method}
  \label{fig:time}
\end{figure}

The basic algorithm ITH requires $4267$ sec in order to find the best ranking vector, as can be seen in Table \ref{tab:timeO}. That is, ITH algorithm requires about $3.3$ sec/image. By employing block randomized SVD to facilitate matrix inversion, the computational time of BR-ITH drops to about $3126$ sec. That is, BR-ITH requires about $2.41$ sec/image, yielding a time reduction approximately $27  \%$ of the time needed by ITH. The $F_{1}$ measure obtained by  BR-ITH  is identical to that of ITH for all ranking positions, as shown in Table \ref{tab:pliakf1}. When the conjugate gradient method is employed within ITH, the computational time reduces to $355$ sec, yielding a gain of about $92  \%$ w.r.t. ITH. That is, CG-ITH requires about $0.27$ sec/image. The resulting $F_{1}$ measure still remains the same with that of ITH.

In the ITH-HWEG we have to optimize not only w.r.t. $\mathbf{f}$ keeping $\mathbf{w}$ fixed, but also $\mathbf{w}$ keeping $\mathbf{f}$ fixed. The basic ITH-HWEG takes $8530$ sec to derive the optimal ranking vector. That is, $1.99$ times longer than the computational time of ITH. By employing the proposed block randomized SVD, the time drops to $4606$ sec, which amounts to $46\%$ reduction in time w.r.t. ITH-HWEG. That is, BR-ITH-HWEG induces a computational cost of about $3.56$ sec/image. The $F_{1}$ measure achieved by BR-ITH-HWEG at certain ranking positions is slightly better than that of the basic algorithm ITH-HWEG (see F1@5, F1@10 in Table  \ref{tab:pliakf1}). By employing the conjugate gradient within ITH-HWEG, the computational time drops to $727$ sec, i.e., to approximately $8\%$ of the computational time required by ITH-HWEG (i.e. reduction $92\%$). The computational reduction does not at all affect the $F_{1}$ measure at the four ranking positions listed in Table \ref{tab:pliakf1}. Thus, the conjugate gradient method can be exploited in large datasets and real-time applications for a single query image ($0.56$ sec/image), which might include millions of images for optimizing $\mathbf{f}$ given $\mathbf{w}$. Figure \ref{fig:time} depicts the overall computational time needed by all algorithms for various number of images. ITH, BR-ITH, and CG-ITH optimize $\mathbf{f}$ given a fixed $\mathbf{w}$. ITH-HWEG, BR-ITH-HWEG, and CG-ITH-HWEG run an alternating minimization problem of $\mathbf{f}$ given $\mathbf{w}$ and $\mathbf{w}$ given $\mathbf{f}$. It is seen that the second optimization step of optimizing $\mathbf{w}$ given $\mathbf{f}$ when it is included in the loop improves $F_{1}$ measure, but at the expense of larger computational time than that of ITH. In the latter case, the computational time increases almost linearly with the number of images, which could be prohibitive.

\begin{table}[t]
\caption{Time requirements for the optimization problems (\ref{eq:Busolution}) and (\ref{eq:conjLinearSys}) in sec}
\label{tab:timeO}
\begin{center}
\begin{tabular}{| >{\centering\arraybackslash}m{1in} | >{\centering\arraybackslash}m{0.81in} | >{\centering\arraybackslash}m{0.91in} |}
\hline
Methods&ITH \cite{Pliakos:14tag}&ITH-HWEG \cite{7471862}\\ \hline
 Original &  4267 &  8530 \\ \hline
  Block randomized  &  3126 & 4606 \\ \hline
 Conjugate Gradient &  355 & 727 \\ \hline
\end{tabular}
\end{center}
\end{table}

\begin{table}[htb]
\begin{center}
\caption{$F_{1}$ measure at various ranking positions for various approaches} \label{tab:pliakf1}
\begin{tabular}{| >{\centering\arraybackslash}m{1in}|l|l|l|l|}
\hline
 Methods&$F_{1}@1$&$F_{1}@2$&$F_{1}@5$&$F_{1}@10$\\ \hline
 ITH \cite{Pliakos:14tag} &0.312&0.457&0.530&0.445\\ \hline
 BR-ITH             & 0.312 & 0.456 & 0.531 & 0.444\\ \hline
 CG-ITH            &0.312  &  0.456 & 0.530 & 0.445\\ \hline
  ITH-HWEG \cite{7471862} &0.425 & 0.682 & 0.753 & 0.558\\ \hline
 BR-ITH-HWEG             &0.425 & 0.674 & 0.756 & 0.562\\ \hline
 CG-ITH-HWEG             &0.427 & 0.680 & 0.752 & 0.557\\ \hline
\end{tabular}
\end{center}
\end{table}


\section{Conclusions and Future Work}
\label{sec:concl}

Two different approaches for optimizing the ranking vector in hypergraph learning have been proposed. These methods are block randomized SVD for matrix inversion and conjugate gradient for solving a set of linear equations both related to optimizing $\mathbf{f}$ given fixed hyperedge weights $\mathbf{w}$. It was shown that both approaches yield computational time savings without affecting the $F_{1}$ measure at various ranking positions. The most prominent shortcomings have been obtained by employing the conjugate gradient method in ITH and ITH-HWEG. The disclosed results are promising and motivate further research towards accommodating bigger image datasets than that considered.




\bibliographystyle{IEEEbib}
\bibliography{bibliography}

\begin{thebibliography}{10}

\bibitem{Kotropoulos:15}
C.~Kotropoulos,
\newblock ``Multimedia social search based on hypergraph learning,''
\newblock in {\em Graph-Based Social Media Analysis}, I.~Pitas, Ed., vol.~39,
  pp. 215--273. CRC Press, 2016.

\bibitem{halko2011finding}
N.~Halko, P.~Martinsson, and J.~A Tropp,
\newblock ``Finding structure with randomness: Probabilistic algorithms for
  constructing approximate matrix decompositions,''
\newblock {\em SIAM Review}, vol. 53, no. 2, pp. 217--288, 2011.

\bibitem{Hatamirad2018}
S.~Hatamirad and M.~M. Pedram,
\newblock ``Low-rank approximation of large-scale matrices via randomized
  methods,''
\newblock {\em J. of Supercomputing}, vol. 74, no. 2, pp. 830--844, 2018.

\bibitem{martinsson2011randomized}
P.~Martinsson, V.~Rokhlin, and M.~Tygert,
\newblock ``A randomized algorithm for the decomposition of matrices,''
\newblock {\em Applied and Computat. Harmonic Analysis}, vol. 30, no. 1, pp.
  47--68, 2011.

\bibitem{liberty2007randomized}
E.~Liberty, F.~Woolfe, P.~Martinsson, V.~Rokhlin, and M.~Tygert,
\newblock ``Randomized algorithms for the low-rank approximation of matrices,''
\newblock {\em Proc. of the National Academy of Sciences}, vol. 104, no. 51,
  pp. 20167--20172, 2007.

\bibitem{teng2018fast}
D.~Teng and D.~Chu,
\newblock ``A fast frequent directions algorithm for low rank approximation,''
\newblock {\em IEEE Trans. Pattern Analysis and Machine Intelligence}, 2018, to
  appear.

\bibitem{grigori2018low}
L.~Grigori, S.~Cayrols, and J.~W Demmel,
\newblock ``Low rank approximation of a sparse matrix based on lu factorization
  with column and row tournament pivoting,''
\newblock {\em SIAM J. Scientific Computing}, vol. 40, no. 2, pp. 181--209,
  2018.

\bibitem{oh2018fast}
T.~Oh, Y.~Matsushita, Y.~Tai, and I.~Kweon,
\newblock ``Fast randomized singular value thresholding for low-rank
  optimization,''
\newblock {\em IEEE Trans. Pattern Analysis and Machine Intelligence}, vol. 40,
  no. 2, pp. 376--391, 2018.

\bibitem{Yu:12a}
Z.~Yu, S.~Tang, Y.~Zhang, and J.~Shao,
\newblock ``Image ranking via attribute boosted hypergraph,''
\newblock in {\em Proc. 13th Pacific-Rim Conf. Advances Multimedia Inf.
  Process.}, 2012, pp. 779--789.

\bibitem{Xu:12}
J.~Xu, V.~Singh, Z.~Guan, and B.~S. Manjunath,
\newblock ``Unified hypergraph for image ranking in a multimodal context,''
\newblock in {\em Proc. IEEE Int. Conf. Acoustics, Speech, and Signal
  Process.}, 2012, pp. 2333--2336.

\bibitem{Gao:13}
Y.~Gao, M.~Wang, Z.~J. Zha, J.~Shen, X.~Li, and X.~Wu,
\newblock ``Visual-textual joint relevance learning for tag-based social image
  search,''
\newblock {\em IEEE Trans. Image Process.}, vol. 22, no. 1, pp. 363--376, 2013.

\bibitem{Pliakos:15}
K.~Pliakos and C.~Kotropoulos,
\newblock ``Weight estimation in hypergraph learning,''
\newblock in {\em Proc. IEEE Int. Conf. Acoustics, Speech, and Signal
  Process.}, 2015, pp. 1161--1165.

\bibitem{7471862}
A.~Chasapi, C.~Kotropoulos, and K.~Pliakos,
\newblock ``Adaptive algorithms for hypergraph learning,''
\newblock in {\em Proc. IEEE Int. Conf. Acoustics, Speech and Signal Process.},
  2016, pp. 1179--1183.

\bibitem{hackbusch1999sparse}
W.~Hackbusch,
\newblock {\em A Sparse Matrix Arithmetic Based on H-Matrices. Part I:
  Introduction to H-Matrices}, vol.~62,
\newblock Springer, 1999.

\bibitem{Pliakos:14tag}
K.~Pliakos and C.~Kotropoulos,
\newblock ``Simultaneous image tagging and geo-location prediction within
  hypergraph ranking framework,''
\newblock in {\em Proc. IEEE Int. Conf. Acoustics, Speech, and Signal
  Process.}, 2014, pp. 6944--6948.

\bibitem{Zhou:06}
D.~Zhou, J.~Huang, and B.~Sch{\"o}lkopf,
\newblock ``Learning with hypergraphs: Clustering, classification, and
  embedding,''
\newblock in {\em Advances Neural Inf. Process. Systems}, 2007, vol.~19, pp.
  1601--1608.

\bibitem{Agarwal:06}
S.~Agarwal, K.~Branson, and S.~Belongie,
\newblock ``Higher order learning with graphs,''
\newblock in {\em Proc. 23rd Int. Conf. Machine Learning}, 2006, pp. 17--24.

\bibitem{Zhou:03}
D.~Zhou, O.~Bousquet, T.~N. Lal, J.~Weston, and B.~Sch{\"o}lkopf,
\newblock ``Learning with local and global consistency,''
\newblock in {\em Advances Neural Inf. Process. Systems}, 2004, vol.~16, pp.
  321--328.

\bibitem{Bu:10}
J.~Bu, S.~Tan, C.~Chen, C.~Wang, H.~Wu, Z.~Lijun, and X.~He,
\newblock ``Music recommendation by unified hypergraph: {Combining} social
  media information and music content,''
\newblock in {\em Proc. ACM Conf. Multim.}, 2010, pp. 391--400.

\bibitem{barrett1994templates}
R.~Barrett, M.~Berry, T.~F. Chan, J.~Demmel, J.~Donato, J.~Dongarra,
  V.~Eijkhout, R.~Pozo, C.~Romine, and H.~Van der Vorst,
\newblock {\em Templates for the Solution of Linear Systems: Building Blocks
  for Iterative Methods},
\newblock SIAM, Philadelphia, 1994.

\end{thebibliography}

\end{document}